# An Ontology-based Adaptive Personalized E-learning System, Assisted by Software Agents on Cloud Storage

Monika Rani, Riju Nayak and O.P. Vyas
IIIT Allahabad
*monikarani1988@gmail.com

*Abstract*—E-Learning and online education have made great strides in the recent past. It has moved from a knowledge transfer model to a highly intellect, swift and interactive proposition capable of advanced decision-making abilities. Two challenges have been observed during the exploration of recent developments in e-learning. Firstly, to incorporate e-learning systems effectively in the evolving semantic web environment and secondly, to realize adaptive personalization according to the learner's changing behaviour. An ontology-driven system has proposed to implement the Felder-Silverman learning style model in addition to the learning contents, to validate its integration with the semantic web environment. Software agents are employed to monitor the learner's actual learning style and modify them accordingly. The learner's learning style and their modifications are made within the proposed e-learning system. Cloud storage is used as the primary back-end in order to maintain the ontology, databases and other required server resources. To verify the system, comparisons are made between the information presented and adaptive learning styles of the learner along with actions of agents according to learners' behaviour. Finally, various conclusions are drawn by exploring the learner's behavior in an adaptive environment for the proposed e-learning system.

Keywords—e-learning, ontology, semantic web, cloud storage, software agents, adaptive.

## 1. INTRODUCTION

The emerging semantic web needs to develop an e-learning system which focuses on personalized and adaptive learning style of learners rather than just content delivery. Current E-learning system of read/write web (web 2.0) is facing some challenges to meet the requirements of semantic web (Web 3.0). Some issues of current e-learning system are to manage huge continuous growing e-learning content on the internet, searching an appropriate e-learning content as per the learner's requirement, represent knowledge in machine readable format with reasoning capability and also to allow reuse of e-learning material. All these issues are addressed by using ontologies for storing e-learning content and building an e-learning application for the semantic web. The ontology presents the course taxonomies in an unambiguous format which is its main resolution. Machine–readability and parsing capabilities of ontology makes it ideal for collaborative purposes. The knowledge base can be shared with other applications of similar intent. In our proposed e-learning system, the detail of personalization is stored appropriately in ontology, based on the Felder-Silverman model [1] and dynamic changes are notified by JADE agents. Ontology provides personalized e-learning content as the learner's requirements change dynamically and the agents capture these changes in learning style and store this information in the ontology. The agents collaborate to thus provide accountability for adaptive learning. To store ontologies we require an expanded and secure environment, thus the entire system is deployed on DigitalOcean's remote cloud host. Cloud can store incremental e-learning content and also provides security by preventing unauthorized access of e-learning content. By overcoming these issues we can develop an effective and enlightening ontology driven personalized and adaptive e-learning system.

The rest of this paper is structured into 6 sections; initially we focus on introduction of paper in section 1. In Section 2, Foundation provides an overview of technologies used to propose our system, like, e-learning, ontologies, cloud computing for storing ontology and multi-agents architecture for interaction among agents. In Section 3, we explain the methodology details with the specification related to the actual technologies engaged on the basis of the foundation of the system. Methodology mentions the technologies used like Felder-Silverman learning model, ontology building tools and languages, DigitalOcean's cloud hosting and multi-agent architecture for development of proposed ontology driven adaptive & personalized e-learning system. Section 4 shows experimental results, which will affirm the content provided to the learner by the system, by imitating to the learner's learning style. It also mentions the agents' actions and their impact on the adaptive nature of personalization realized by the system. Followed by section 5, in which Learner's dimension, Instructor dimension, Course dimension, Technology dimension and the Design dimension are used to evaluate the effectiveness of the proposed e-learning system. Finally, section 6 draws conclusion and future research opportunities in the current e-learning scenario.

## 2. FOUNDATIONS

### 2.1 E-learning and readiness

E-learning has gradually emerged as one of the most frequently used technologies in the modern era. The importance of e-learning is highlighted through emphasizing learning techniques as well as patterns. Hisham et al. has briefly defined it as a learning platform that utilizes information and communication technologies as well as electronic media. They also implied a number of alternative terms for e-learning such as technology enhanced learning, computer-based training, online-education, and others [2]. This definition quite immensely generalizes the utility of e-learning, which is of high importance, as the scope and approach of constructing an e-learning system is heterogeneous. Focus on a particular design of a system may vary completely from another design and heterogeneity in it leads to segregation of research areas within e-learning. Various segregations require different approaches to actualize the desired system.

A few studies have been reviewed in order to concisely comprehend the readiness of e-learning as a field. A study in 2004 conducted by the U.S. Coast Guard (USCG) focused on the validation of e-learning readiness and was achieved internally via checking the consistency of objects assigned to the development of self-assessment. Data obtained from it, was later employed as a guide for better enhancements that seemed fit for the development of numerous instruments working towards the cause. Respondents included members from the USCG within the age range of 17 to 34. Despite the study focusing on online learning, the respondents didn't have to actively use any online courses. The assessment of results confirms the potential in terms of validation and consistency, and it also shows indication of a good prediction tool in determining e-learning performance [3].

In 2005, Directors of Human Resource Department of, several companies in Turkey started an initiative to assess e-learning readiness in emerging countries. Top 100 companies in Turkey have been selected to become a part of a survey, by the Istanbul Chamber of Industry (ICI), from it's of 500 Major Industrial Enterprises of Turkey, published in 2001. While achieving a precise review of their e-learning readiness, they arrived at a conclusion to develop the companies' HR structure before proceeding with the integration of the online courses [4].

In 2006, a study focused to unravel the readiness among the teachers of an institute, rather than its students, was conducted by the academic staffs and deans of the International Islamic University Malaysia. The study underlined two factors, which played a crucial role in determining their readiness, e-learning training and confidence. However, it was suggested that their improvement hinged upon the infrastructure of the institute. The study also concurred that gender did not play a factor among its respondents [5].

In 2008, a study was conducted to review the feasibility in the health department, which was done by the nurses in Flemish hospitals in Belgium. The analysis also highlighted the necessity of training along with determining the importance of strict protocols involving work hours. It also emphasized the importance of transparency between communication involving the developers and people in charge of hospital policies [6].

In 2011, the focus of assessing e-learning readiness was to candidly determine its acceptance among students of different levels of proficiency in a subject. A group of undergraduate students studying English as a Foreign Language were selected from the King Khalid University in Saudi Arabia as respondents. The study showed a complete acceptance in e-learning integration in their environment [7].

In 2013, a study was conducted to determine the readiness of PhD scholars in the Christian University of Thailand. Students were selected from various years into their research and the aspects taken into account, while quantifying their acceptance were technology access, online audio/video, importance to success, internet discussion, online skill and relationship, and motivation. As a whole, the study uncovered a great extent of e-learning readiness, wherein, technology access proved to be the most promising aspect while motivation was theleast. The difference in demographics, according to their year in research or gender had no significance at the readiness level [8].

In 2015, Satpute et al. reviewed several prototypes engaged for educational needs and compared their usability to find the advantages of using Augmented Reality (AR). Web 2.0 tools were also examined to understand the combined use of the two technologies. They concluded by assertingbetter results in educational achievements come through by combining technologies.

AR enhances immersion and engagement, whereas web 2.0 supports social interaction and collaboration [9].

*2.2 Ontology – Type, Specification Languages, Development Methodologies and various application areas:*

The learner personalization details as well as the taxonomy of learning resources will be maintained in ontology. There have been many attempts to define an ontology, though all of them have described the same concept but from different perspectives. However, in 1998, Studer et al. made an attempt to define the term while keeping in consideration, all the contemporary perspectives and stated that – "An ontology is a formal, explicit specification of a shared conceptualization" [10], the definition which needs a thorough explanation to decipher a comprehensive understanding. The word formal implies that the knowledge or content represented by the ontology is stored in a format understood by computers, which makes parsing the content, trivial. In the paper, DL Query[1] is used as the query language to parse the data. The explicit specification means that no relationships or concepts depicted by the ontology can be assumed or is implicit. Every property and relationship must be listed in its entirety, with none left to be assumed, which could result in multiple inferences. Finally, shared conceptualization reflects knowledge and its constituent conception to have a definition entirely agreed upon. This entails that the content represented by the ontology is universally accepted and only has a single perspective and context to understand. In the system, domain ontology is used, as the personalization to be demonstrated is done sufficiently through a localized batch of concepts pertaining to a specific domain. Elaboration is needed for the development of domain ontology to provide personalized e-learning.

*2.2.1 Ontology definition from the prospective of e-learning an application area:*

- Having a formal representation of knowledge is helpful in interoperability within heterogeneous e-learning environments.
- Explicit specification goes towards the enhancement of exhaustive learning by not making assumptions on the implicit nature of the information or the learner's style.
- Shared conceptualization ensures that the knowledge being stored and used, has no ambiguity or over its definition.

---

[1]DL Query, Protégé wiki -
http://protegewiki.stanford.edu/wiki/DLQueryTab

Elaboration is needed for the development of domain ontology to provide personalized e-learning. There have been few attempts at creating domain ontologies to meet the e-learning demands, even though its importance is substantial.There are multiple aspects to be considered while classifying ontologies. These aspects may be characterized according to their formality, granularity, computational capability and generality.

*2.2.2 Types of ontology in accordance to the generality will be explained as:*

- *Top-level ontologies* [11] also referred to as upper ontologies or foundational ontologies are domain-independent, high level ontologies. Examples include overly generalized, cross domain ontologies explaining general concepts such as Time, Space and others.
- *Mid-level ontologies*, also known as utility ontologies, behave as a channel between top-level ontologies and domain ontologies. Their purpose is similar to that of software libraries in object-oriented programs.
- *Task ontologies* are developed in order to store content relevant to a specific task, like, presenting fundamental concepts related to an overly general activity or task.
- *Domain ontologies* specify concepts, their properties and relationships pertaining to a specific domain of interest. Example ontologies showing principal concepts which relate to a generic domain. Therefore, the scope needs to be very distinctly specified.
- *Application ontologies* are created for the purpose of aiding specific applications. They generally are the combination of domain and task ontologies. For instance, it includes the most specialized ontologies which are application specific, focusing on a definitive task or domain.

*2.2.3 Ontology languages:*

In this present work, various web ontology languages are explained as the research is focused on the semantic web. A few web ontology languages are OIL, DAML+OIL, SHOE, XOL and OWL. Interoperability is ensured in the web environment, as these languages are based on the web standards XML and RDF.

- *Extended Markup Language (XML)* is a markup language which tries to segregate web content from web presentation. A major drawback is its lack of semantics, although it's widely used as the web standard to represent information [12].
- *Resource Description Framework (RDF)* is a W3C standard used to represent web resources. A statement in RDF is called a triple which consists of a subject, predicate and object. A triple can be imagined as a directed link between two nodes, which can be modeled as the subject and the object, whereas the predicate acts as a directed link which is from the subject to the object. The purpose of RDF is to allow exchange of machine-understandable information, mainly on the web [13].

- *Ontology Interface Layer (OIL)* was developed during the On-To-Knowledge project. It is established on frame-based languages, description logics and web standards. Its purpose is for both representing and exchanging ontologies [14].

- *DAML+OIL* are the consequence of an effort to merge DARPA Agent Markup Language (DAML) and OIL. DAML+OIL show more efficiency than OIL due to increase in features from description logics. However, due to the exclusion of several frame-based features, usability with frame-based tools became limited [15].

- *An XML-based Ontology Exchange Language (XOL)* is designed as a framework to exchange ontology definitions [16].

- *Simple HTML Ontology Language (SHOE)* extends and allows HTML pages to incorporate machine-readable semantic knowledge [17].

- *Web Ontology Language (OWL)* is a standard for representing ontologies on the semantic web. Web-Ontology (WebOnt) Working Group developed it in 2001. It soon became a W3C recommendation in 2004. OWL provides developers with a superior power to express semantics, and to allow automated reasoners to derive knowledge and to carry out logical inferences [18].

*2.2.4. Building ontology:* Methodologies for the development of ontologies can be traced back to the time of development of the Cyc ontology, during which Cyc developers published their experiences [19]. Subsequently, experience in developing the Enterprise Ontology [20] and the TOVE (Toronto Virtual Enterprise) [21] ontology was also reported. After these proposals, a series of ontology development methodologies were presented, including KACTUS [22], METHONTOLOGY [23], Sensus [24], On-To-Knowledge [25] and CO4 [26].

In this article, the focus is on the method proposed in the KACTUS project, in which, the ontology is built in a bottom-up manner from a knowledge base (KB) application. It was achieved through abstraction which selects a KB for a specific application, and when the need arises to create an ontology for a similar application, the first KB abstracted represents an ontology for both applications.

*2.2.5 Ontology Applications in various fields:*

- Ontology based Question and Answering: Ontology plays an important role in the development of knowledge based system which describes semantic relationships among entities. These relationships described with the help of ontology are able to reach accurate answers to a user's question. Question Answering systems; improve if the emphasis is on the semantic analysis of literal terms in a user's query rather than the syntax analysis. The Fuzzy Ontology plays a vital role in understanding such ambiguous user questions. Fuzzy Ontology can help in understanding Semantic relationships by applying Fuzzy logic (Fuzzy Type 1 and Fuzzy Type 2) to deal with vagueness [27]. Fuzzy type-1 can deal with Crisp membership, whereas Fuzzy type-2 deals with Fuzzy membership.

- Ontology based decision support system: Ontology can be used along with various methods to build intelligent discussion support system. Ontology-supported case-based reasoning (OS-CBR) method [28].

- Ontology based E-commerce services: Ontologies provide web services through various heterogeneous domains by using Agent technologies. In an environment Agent allows us to access, retrieve and process relevant content from various domain ontologies. Ontology integrates with the Multi - agent platform to improve e-commerce application for B2B and B2C. For example, in real world scenarios if user wants to retrieve information like flight booking, hotel reservation, banking transaction, etc. information from the heterogeneous domain ontology in single click is hard to obtain results, as we first need to integrate these domains. The main goal of this research is to develop a real time system with the help of ontologies to support systems like railways, flight ticket booking domain, hotel booking domain, banking transaction system, information retrieval system etc. by communicating among agents having domain ontology knowledge. Fuzzy ontology with multi-agent platform system (FOMAS) to give a proposal to automate the personalized example for flight ticket booking domain. Fuzzy type 2 is helpful for retrieval with multi-agent platform system (T2FOMAS). When we consider security aspect of such system called secure type fuzzy ontology multi agent system (ST2FOMAS) [29].

- Ontology based Recommender system: Wu et al. describe Fuzzy set technique can be used to express user preference for recommendation of items, e-books in business like e-service [30] and e-learning.
- In 2013, Sarmacakulaet al. indicate overlapping point between e-learning and knowledge management to provide personalized course content. The merging of various dimensions like student personality using Myers-Briggs learning styles, course content is described in the form of IEEE Learning objects (LO), TECHNOLOGY USING FIAP device ontology and at knowledge level using taxonomy to provide personalized content [31-52].

*2.3 Cloud Computing*

In the proposed architecture cloud storage is used as its primary back-end. The cloud is where the proposed system is deployed along with its complementing learner/user database, ontologies such as the *user.owl* and *course.owl*, and the relevant learning resource for the courses. The advantages of using cloud storage can be directly derived from many of its generic advantages.

- The cost of backing up data on a cloud is incremental, therefore saving large costs of physical servers.
- Cloud storage is said to be invisible, implying that there isn't any prerequisite for a physical space to store the servers required for the proposed system.
- Security becomes unquestionable, as it is enforced via DSA public key encryption with the remote terminal.
- Many cloud services provide several APIs to work with by default, therefore increasing automation from the beginning.
- Cloud Storage has extensive features in its online GUI to further ease backing up and downloading.
- Accessibility of cloud storage is by default extended to multiple platforms such as mobile phones and tablets along with real-time syncing.

Salesforce.com marked the advent of cloud computing. The service utilized the internet to grant business applications in 1999. Software-as-a-Service (SaaS) is what this is called now. In 2002, Amazon Web Services (Amazon Mechanical Turk) was introduced, following which, in 2006, Elastic Compute Cloud (EC2) was launched as a commercial web service providing computing capacity. Amazon is credited with pioneering in pay-as-you-go services and the two services established a firm foothold in cloud computing. With the introduction of Web 2.0, and Google, others got into the business of providing browser-based enterprise applications. Cloud computing became apparent as the next step in evolution and took over computing models such as cluster, grid, utility, distribution and services computing [32].

Although there's limited clarity on the emerging trends in cloud, the following two topics noticed: Technological facets such as "elasticity", "multitenant", "resource pooling", "computing", and "virtualization" and Commercial faucets such as "self-service on demand", "SaaS", and "pay-per-use" [33].

Cloud computing has received considerable attention at both firm and industry levels from 2007. The cloud computing environment modifies the role of the computing stakeholders and also demands, regulatory compliance with the infrastructure pertaining to the location of the service provider [34]. The cloud provides flexible IT resources and has completely changed the way they are provided. Cloud computing on-demand resources allow IT organizations to scale quickly as the need arises which is required by the ever increasing business needs.

E. Aljenaa et al. propose an e-learning framework to store rapidly developing e-learning resource on cloud as due to its scalability, thus providing E-learning as a service (EaaS) [35].

*2.4 Multi-agent architecture*

A multi-agent architecture is used in the proposed system and a few of its properties are listed below, which makes it advantages clear.

- The primary advantage of its usage can be narrowed down to the main agenda of the system, which is personalization.
- The multiple agents used can be categorized as intelligent software agents which are autonomous, specialized in their roles and are persistent entities.
- Their multi-agent environment adds the adaptability factor to personalization by monitoring learner's activities to modify initial learning style assessment from the Felder-Silvermanmodel [36].
- Agents differ from a normal program module in their resistance towards the environment.
- Software agents maintain their own code, but are dormant in execution until triggered by the environment.
- Their power lies in the fact that they can be modified without affecting its environment which includes collaborating agents.

The relevance of personalization became clear through the need for more specific materials according to learners' preferences which in turn improved learners' performance. Novel methodologies as well as proposed framework were introduced to achieve personalization, with encouraging results indicating enhanced learner performance [37].

Deploying intelligent tutor in online education emerged in the 90s of the last century. Sherman Alpert and his colleagues conducted a research in 1999 on the shift of using independent Intelligent Tutoring System (ITS) compared to those operating on the World Wide Web, showing both, architecture and features of the system, support students' problem solving activities [38].

Another research conducted by Marcia Mitchell proposed a framework for an intelligent tutoring system that support Distance Learning (DL), which is high level software based tutoring that has the ability to encompass a wide variety of current DL technologies in a single session [39].

In 2012 a new trend commenced, which focused on deploying agents in learning, applications based on mobile games. Two studies were conducted autonomously, one in English [40] and another in Japanese Kanji. The results confirmed acceptance towards the use of the frameworks proposed, which were analogous to the multimedia world [41].

Dung et al. propose an architecture in 2011 which is to represent domain ontology for e-learning course content and agents interact on it to provide personalized e-learning content to learners [42]. Subsequently, in 2013 Bokhari et al. propose an architecture for interactive multi-agent based learning system for distance education [43].

In 2015, a study applied Multi-Agent Systems (MSA) in several research projects and software agent based projects. With the aid of MAS methodology, a tool was developed, known as the Agent tool, to understand the role of MAS in the design of distributed software systems [44].

## 3. METHODOLOGY

A general architecture of the system is presented in Fig. 1. It stresses on the key technologies to be used in the proposed system. A web-based application is built and deployed on DigitalOcean's remote cloud host. All of its required resources are stored in the cloud storage as well. The domain ontology stored in the cloud will be referred in the application to query and present required information to the learner. A simple database, which is provided by the MySQL Server running on the cloud host, is used for authentication of the system. The actual learning material will be available to the application via linking their cloud URL in the domain ontology attributes. A log file is maintained in the front end to maintain learner activities relevant to modification in their learning style. The proposed system is likely to function as expected, without agents.

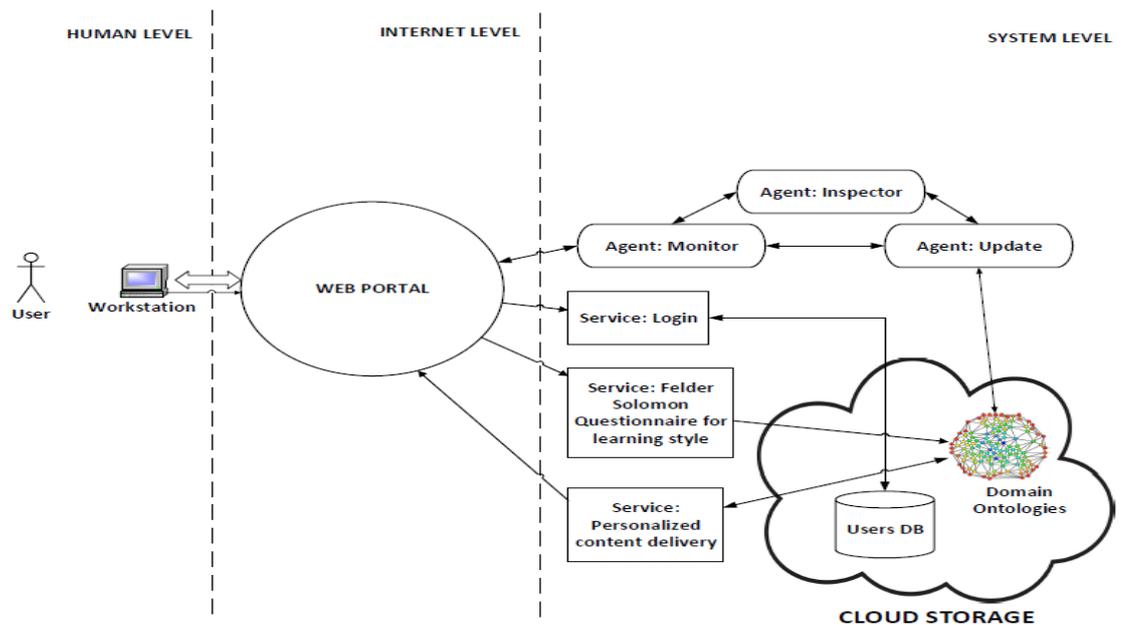

Fig.1. Architecture of Proposed System

However, agents are crucial to the adaptive nature of the personalization. Domain ontologies include the *user.owl* which stores the learner's personalization details. The *course.owl* stores taxonomy of the course outline which is eventually linked to its relevant learning material. Agents utilize the maintained log file to provide added personalization on top of the established learning style derived from the Felder-Solomon questionnaire [45].

*3.1 Felder-Silverman Learning Model*

In 1988, Engineering Education published the Felder-Silverman model as an article called "Learning and Teaching styles in Engineering Education" to offer insights about teaching and learning. The study is based on Silverman's competence in educational psychology and Felder's involvement during engineering days. In the next 10 years, their study winded up being the most frequently cited paper in articles published in the Journal of Engineering Education. In recent works, it has been used as the most comprehensive, yet a simple learning model referred for personalization in e-learning [36].

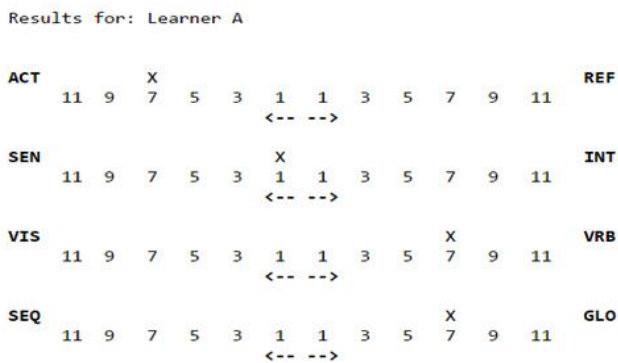

Fig.2. Results of the ILS questionnaire

An overview of several learning style models has been inspected, including Myers Briggs, Gagné's Theory of Learning Styles, Kolb Learning Style Learning Style Inventory, The Ned Herrmann Whole Brain Dominance Theory, and The Gregorc Style Delineator [46]. The analysis led to an understanding that the Felder-Silverman learning style model had measurable dimensions that could directly relate to e-learning aspects. Although the other learning models had persuasive measures and well-founded methodologies in its determination, the Felder-Silverman learning model has been deduced to have learning style indexes that could be realized accurately through the presented information [47].

[2]http://www.engr.ncsu.edu/learningstyles/ilsweb.html

The learning style has been demonstrated for two reasons; firstly to capture the most important learning style differences among engineering students and secondly, to provide a good foundation for engineering instructors to design a teaching approach that would address the learning needs of all students. The second purpose has been mapped to the personalization design of the system. Student's learning styles, according to the Felder-Silverman model, are classified into one of these categories in each of the following four learning dimensions: sensing or intuitive, visual or verbal, active or reflexive and sequential or global.

The Felder-Solomon Index of Learning Styles (ILS)[9] is a questionnaire developed by Richard Felder and Barbara Solomon in 1991. The learner's preferred dimension in the learning style model is determined by the questionnaire. A total of 44 questionsareasked with a compulsory answer. 11 questions are asked for each dimension. Each question has a possible 'a' or 'b' answer that correlates to either one of the categories related to the dimension – for example – the active or reflexive dimension. The 'b' answers are subtracted from the 'a' answers to obtain a score which was an odd number between -11 and 11.

Fig. 2, depicts the results obtained following the questionnaire constructed by Felder and Solomon. It shows all 4 dimensions categorized by their pole characteristics over a scale ranging from -11 to 11. The 'X' on a number indicates the learner's value for that particular dimension.

*3.2 Ontology – Methodology, Tools & Languages*

*Methodology:* The methodology used is a derivation of the *KACTUS* project, in which, the ontology is developed from a knowledge base (KB) application, by abstracting the content to a degree which is satisfactory. In the proposed approach, the KB is textbooks or online learning resources from which the ontology is developed in a bottom up manner.

- *user.owl* – This ontology contains learner/user details such as id, name, courses taken, learning styles and such shown as Fig. 3. This ontology is being directly linked to the authentication process.

- *course.owl* – Learning material and its module of the course is stored in ontology as shown in Fig. 4. Textbooks and learning materials contain the highest degree of details. Starting from that level, the ontology is created by abstracting the details,to reach a level of classification that allows creating a satisfactory taxonomy.

*Tools:* Protégé Ontology Editor[3] is used to develop the ontology. The editor provides a GUI to achieve the same as shown in Fig. 5. It has a highly pluggable architecture, providing easy expansion of various utilities. Protégé is open source software that was developed at Stanford University in association with the University of Manchester, and is made accessible under the Mozilla Public License 1.1. *OWL API 3.5.1* and *HermiT 1.3.8* APIs used in Java to implement ontologies via Programming and *HermiT* is used as a reasoner to verify created ontologies. The *HermiT* is run to make sure there are no inconsistencies inside the ontology.

*Languages:* OWL (Web Ontology Language) is used to specify the ontology in RDF/XML format. For querying, DL Query 2.0.2 was used which is a representative of the Manchester Syntax[4] of OWL. DL Query is used as the query language to parse the data. Example Fig. 6 shows the list of students attending a lecture.

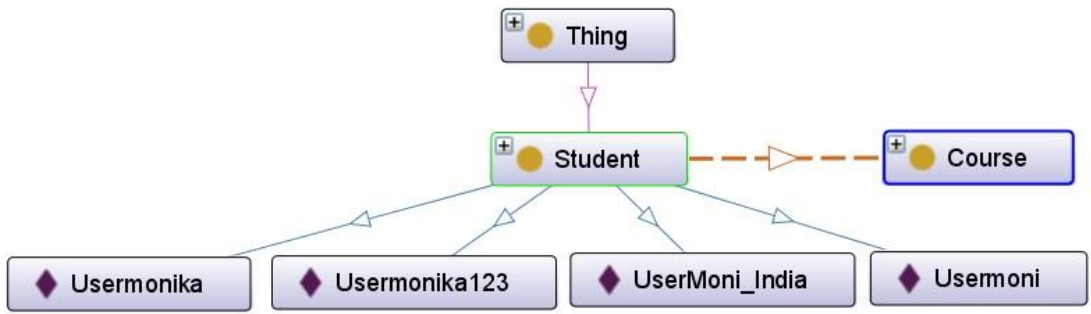

Fig.3. Learner/ user ontology (user. owl)

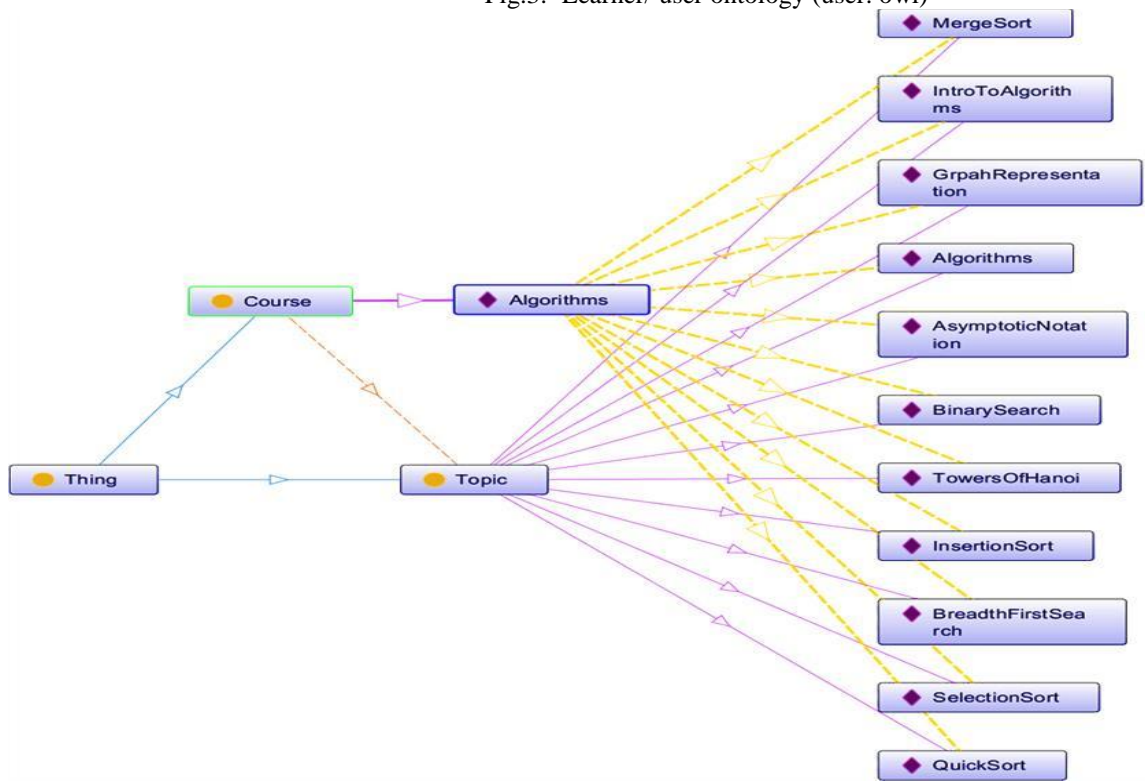

Fig.4. Course ontology abstracted from a KB (course.owl)

---

[3]http://protege.stanford.edu/
[4]http://protegewiki.stanford.edu/wiki/Manchester_OWL_Syntax

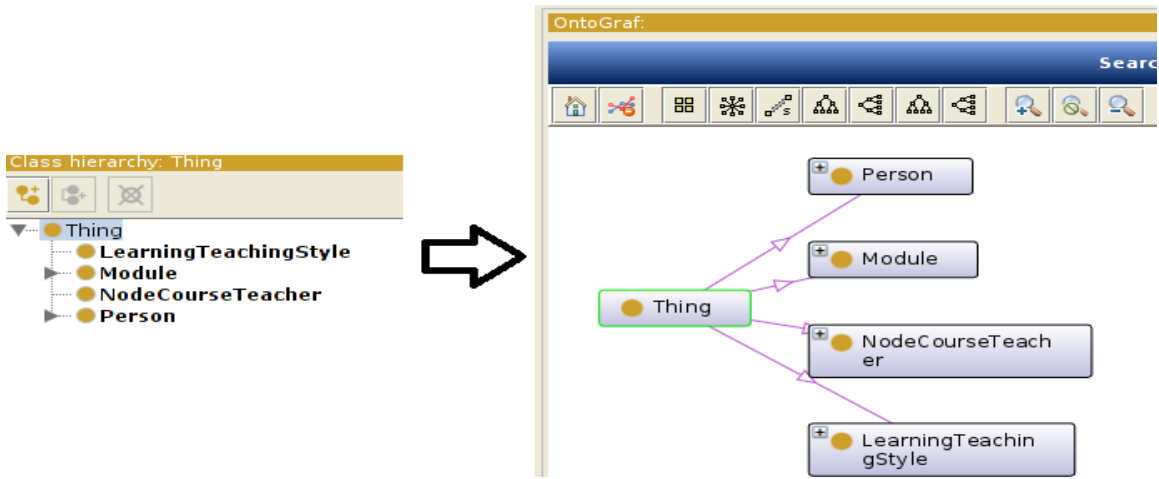

Fig.5. Protégé graphing the created ontology

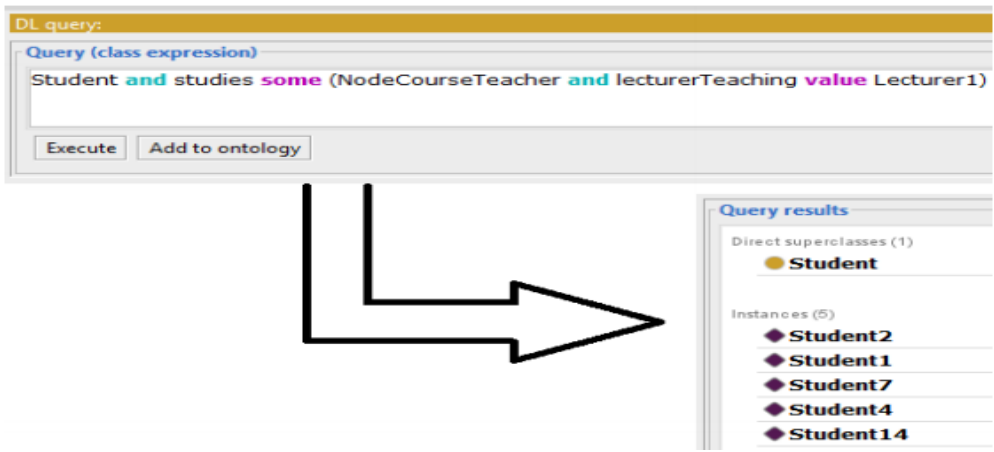

Fig.6. Executed DL Query

Fig.7. Ontologies stored in DigitalOcean Host

### 3.3 DigitalOcean Cloud Hosting

The proposed e-learning system is deployed on the cloud. Its support back-end resources are directly available to it as they are hosted or stored on the cloud as shown in Fig. 7. The agents, however, communicate with the application through the JadeGateway API in Java, which provides a communication channel between the running agents and JSPs or Servlets.

DigitalOcean [48] is an easy-to-use and fast cloud hosting service built for developers. The hoster has its servers based mainly in New York, Amsterdam, and San Francisco. It provides developers with a Virtual Private Server with DNS management. It is a relatively new player into the world of VPS cloud hosting. However, DigitalOcean has expanded as a company tremendously since its founding in 2011. The greatest strides have been made in the last couple of years.

Setting it apart from many different VPS providers, DigitalOcean provides a control panel entirely customizable, designed with the intention of easing. Since we are provided with a Kernel Virtual Machine (KVM) the Linux command line can be accessed directly with superuser privileges, simulating the use of a computer with file system access. The most prominent Linux distros are supplied in the aforementioned control panel during the VPS conception. Selections include 32 bit and 64bit versions of Fedora Desktop, Ubuntu Desktop, Arch Linux, CoreOS, CentOS, and Debian. Utilities such LAMP, WordPress, Ruby on Rails, Docker and Redmine are maintained as "One-click Installs".

- DIGITALOCEAN COMMUNITY: Due to the flexibility of the host, official documentation is unable to cover technology compatibility and usage with the provided KVM. Due to this, DigitalOcean offers a Community. This is essentially a developer-to-developer forum as well as a recognized tutorial on several open source topics. Due to its partnership with Stripe, DigitalOcean is able to sponsor Libscore to freely provide its developer community with free access to analytics on web development tools.
- API V2: DigitalOcean has currently launched its second version of API ('API V2'), which is still in its BETA stage. API V2 is RESTful implying its usage of best practices while creating scalable web services. The API uses OAuth authentication, which allows client applications 'secure delegated access' to resources in the server by acting on behalf of a resource owner. API V2 also supports IPv6.
- LINUX DISTROS: CoreOS and FreeBSD are two unix-based operating systems provided to buyers to work with.
- DATA BACKUP AND RECOVERY: DigitalOcean adds reliability to the stored data by giving the learner two sorts of mechanisms for data backup and recovery. Snapshots can be taken manually of any instance of DigitalOcean, which, however, requires that the VPS be offline for a while. There is also an option to turn on automatic backup as well, which backs up an instance of DigitalOcean periodically.

### 3.4 The Multi-Agent Architecture

The proposed e-learning system being designed is to enhance personalization. Creating a system with a learning model at its foundation provides a great deal of personalization even without agent intervention. However, the running agents are associated with the system to provide adaptive personalization. Before listing out the collaborating agents and their role in context with the system, the JADE technology being used to implement the agents is briefly explained.

JADE (Java Agent Development Framework)[49] is a software framework that eases the development of agents and applications using agents. It is compatible with the FIPA specifications for interoperable intelligent multi-agent systems. Therefore, its goal is to simplify the building of FIPA compliant multi-agent systems, while enhancing its use to contain the features of a FIPA compliant system. To achieve the same, JADE offers the following list of features to the agent programmer:

- Building a FIPA-compliant agent platform, including the AMS (Agent Management System), the DF (Directory Facilitator), and the ACC (Agent Communication Channel). These three agents are the default agents which are activated automatically at the agent platform start-up.

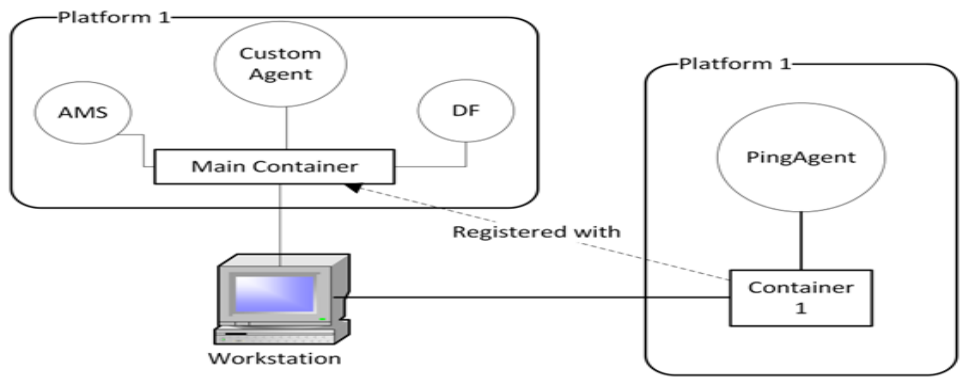

Fig.8.  JADE environment

- JADE provides a distributed agent platform as shown in Fig. 8. A deployed agent platform can be distributed across several hosts connected by a network. A single Java application, and therefore only a single Java Virtual Machine, was executed on each host.
- A single Java thread was used to run agents, and Java events were used for lightweight communications among agents in the same host.
- FIPA-compliant naming services: start-up agents obtain their GUID (Globally Unique Identifier) from the platforms.
- A GUI is provided which can be used to manage several agents running on a platform or multiple platforms as shown in Fig. 9. Generic agents can be deployed on platforms to monitor and log files can be used to describe the agent's activities.

There are cases where a web interface is required with the multiple running JADE agents. The application should be based on JSP and Servlets. JADE offers some utility classes that could help achieve this. The utility comes as an API called the *JadeGateway*. The silver bullet for the utility is the *jade.wrapper.gateway* package in the communication. The package includes these classes:

- *JadeGateway*
- *GatewayAgent*

The system structure of a simple application employing the JadeGateway can be explained via certain keywords and a timeline of events depicting their communication:

- *PingAgent* exists in an agent platform.
- *BlackBoard* is a Java bean created by the servlet and used as a communication channel.
- *GatewayAgent* is created by the servlet too, and it behaves as a dispatcher. It's the main web interface.

The timeline of events to explain their working:

- The browser causes an event generating a POST message.
- The servlet handles it by invoking the send message action.
- A new *BlackBoard* object is created by the invoked action. This Java Bean acts as the communication channel between the *GatewayAgent* and *Servlet*.
- The *GatewayAgent* receives the *BlackBoard* object. From the object, it extracts the recipient and the content of the message. It then forwards the message to the recipient.
- *PingAgent*, which is the recipient, provides its response to the *GatewayAgent*.
- The response from the *PingAgent* is packaged and sent back to the Servlet via the *BlackBoard*.

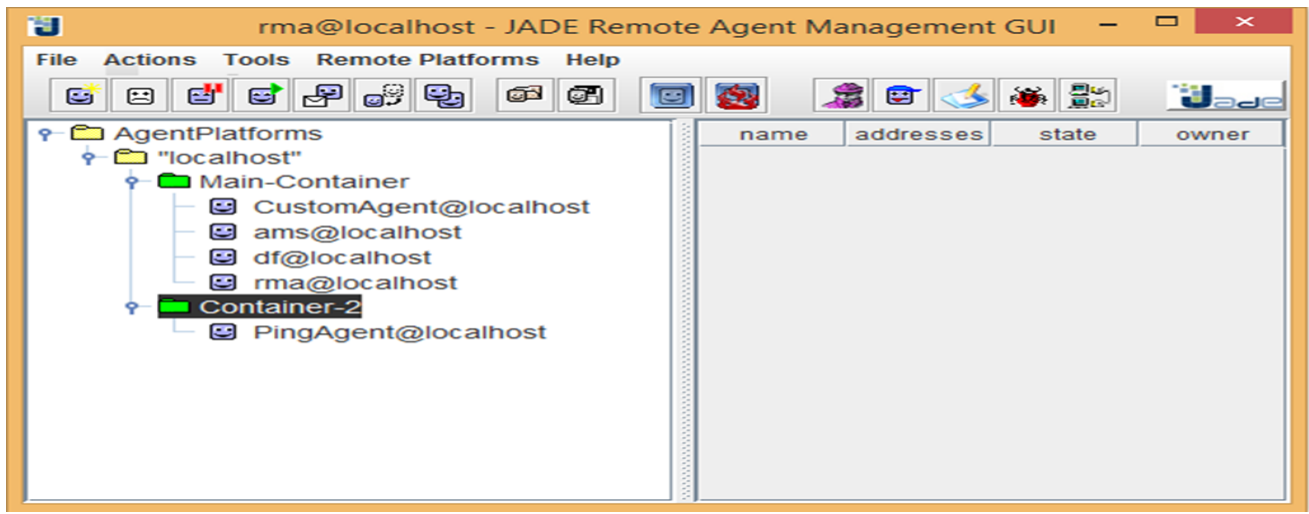

Fig.9. Remote Agent Management GUI is depicting the JADE environment

The JADE platform will be explained via the agent instances running on it. A sequence diagram is depicted in Fig. 10 to illustrate the agent instances as well as their collaboration in order. However, prior to the explanation of the sequence diagram a few words are said about the agents used in the platform for adaptive personalization:

• Agent:Monitor – This agent is contacted by JadeGateway in a behaviour waiting for a message from the Gateway agent which tells the agent of the current user_id in session. The behaviour is cycled in a waiting state. Once it receives the user_id it kills the behaviour and stores the user_id in a static variable. It contains a tickerbehaviour as well, which periodically scans the ontology for changed dimensions. It doesn't start until the agent receives the user_id from the Gateway Agent. In case the changed dimension found are greater than the threshold of above +5 or below -5 the monitor agent informs the update agent.

• Agent:Sniffer – The sniffer agent employs a listener behaviour which intercepts and redirects ACL messages being exchanged. This is a JADE tool, agent which exists to help developers debug their agents to see the interchange of messages and check if they are functioning properly. Here it will also be used to demonstrate the sequence of message passing.

• Agent:Update – This agent employs a cyclic behaviour which is in a waiting state until it receives an INFORM ACL Message from update agent. The inform message contains the changed dimensions and tells the update agent about how much change is required in the dimensions. Once it handles the change, it goes back to its waiting state until it is prompted by the monitor agent again.

The sequence diagram in Fig. 10 is explained below in steps:

• The sniffer agent initiates its ListenerBehaviour to sniff messages across the JADE platform as shown in Fig. 17. As discussed, it is used to notice the sequence of message transfer and lifetime of the agents.

• The website uses JadeGateway's API to initiate Monitor agent's activity by sending it the current user_id in session.

• The monitor agent having received the user_id periodically checks the ontology for dimension changes. If the dimension change is above a certain limit the monitor agent informs the update agent to change dimensions.

• The update agent employs a cyclic behaviour in a waiting state until it receives the INFORM ACL Message from monitor to update the dimensions in ontology. Once it is done, it goes back to its waiting state.

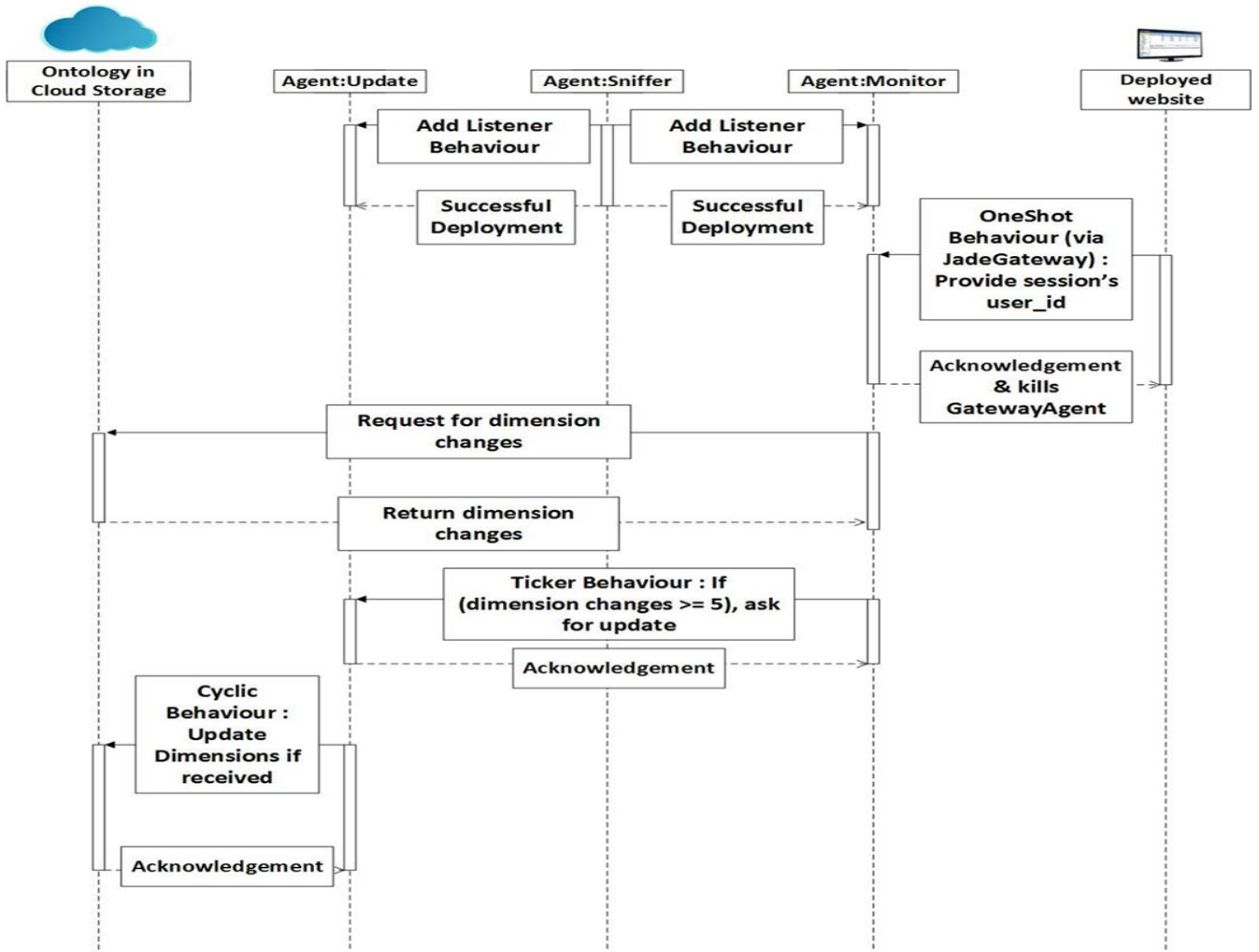

Fig.10.    Sequence diagram of communicating agents

*3.5 Workflow of the proposed Adaptive Personalized E-learning:*

Fig. 11.demonstrates the workflow of the proposed Adaptive Personalized E-learning system. The Modules of the workflow are explained as follows:

- The *course.owl* ontology is created which is responsible for representing the taxonomy of the domain. The majority of the classes in that ontology is either representative of a field or a course. Those classes, pre-dominantly consist of data attributes explicating the URLs of the learning resources pertaining to particular courses.

- The *user.owl* ontology is created to store the learner details and its learning style indexes. Comprehensive research was carried out on several learning models which substantiate learning styles. Following that, the Felder-Silverman learning style model has been exercised, and its complementing Felder-Solomon indexes are stored in the ontology. The ontology also contains information about learner's behaviour which can later be needed to modify an existing style.

The OWL API of Java is used to create the ontologies. A framework for inputting data into the ontologies is created which allows for easy extension of the framework to transform into a GUI based input system. The necessity of such a framework created by the API is to extract data for objects in the ontology and to ease application development in later stages.

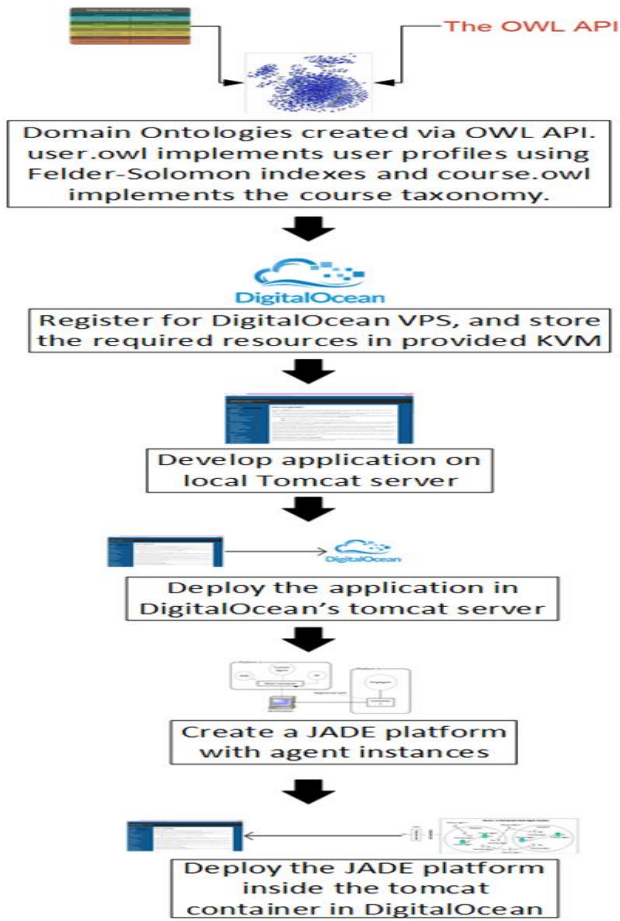

Fig.11. Workflow of Adaptive Personalized E-learning system

- An account must be registered with DigitalOcean. A suitable KVM is chosen according to the application requirement. In our case, a basic system is chosen without any extravagant specifications as our application is purely demonstrative.
- Tomcat and MySQL server are also installed on the registered DigitalOcean host. It's better to deal with the hostname rather than its IP address.
- Tomcat and MySQL were also installed on the local computer on which the app is being developed in order to develop and deploy the application faster. It has been provided for easier and quicker debugging. The complete application can be deployed on the Tomcat container running on the cloud host.
- The JADE environment which allows a FIPA compliant multi-agent system to be created is set up and run alongside the deployed application. The agents are programmed on the localhost itself and tested with a running JADE platform on the localhost before the JADE platform was running on the Tomcat container with the custom agent instances needed for personalization.
- The agents are programmed according to their required application and collaboration needs. These agents were put in containers within the running JADE environment and use the *JadeGateway*communicates with the application.

## 4. .RESULTS AND DISCUSSION

### 4.1 RESULT

Once the learner fills the prerequisite Felder Solomon questionnaire on the basis of provided learning style indexes, the learner is redirected to their personalized learning page accordingly. On the personalized pages, there are hyperlinks which relate to behaviour that might be indicative of changing learning style. In Fig. 12 the 'gallery view' icon is clicked. That triggers a Java program to alter the ontology in a way as shown in the figure. The 'ChangeSG' (corresponding to the change in Sequential-Global dimension) data property is changed from '0' to '-2' to reflect this behaviour. Once this value goes below '-5' or goes above '+5' the actual dimension value is changed this reflects adaptability in learner behaviour.

The results observed in the expected system are shown with the aid of screenshots of the application. Different learning style indexes are shown beside their presented information to highlight the dimensional differences. A single illustration is also shown indicating the communication between the agents responsible for adaptive personalization.

Fig. 13 is indicative of the *Active-Reflexive dimension* of a learner's style. The screenshot on the left of the illustration shows an active learner. As shown in the figure, active learners are provided with challenges regularly, whenever applicable. However, active learners are also provided with a "*Hide Challenges*" option to opt out of regular challenges. If clicked for a certain number of times, the log file used to note a change in the dimension value. The screenshot on the right of the illustration shows a reflexive learner. Regular challenges are hidden from reflexive learners. Similarly, reflexive learners are also delivered with the "*All Challenges*" option to view the challenges provided. If clicked up to a certain time, the log file used to note a change in the dimension value.

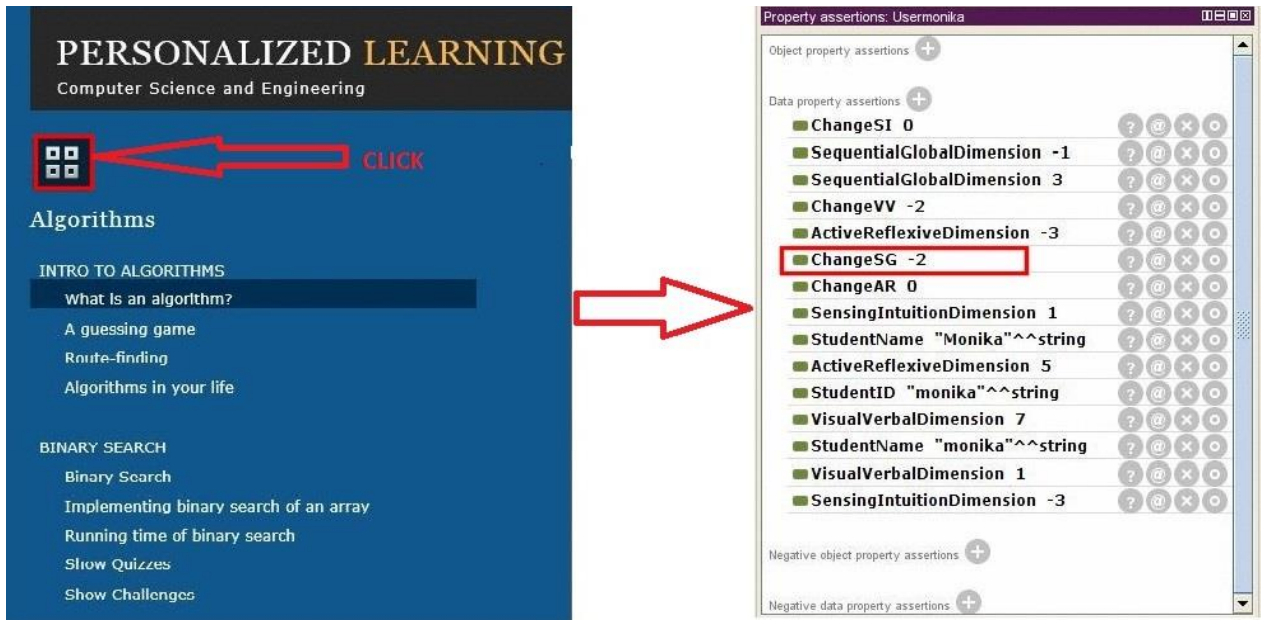

Fig.12.    Learning behaviour observed in ontology

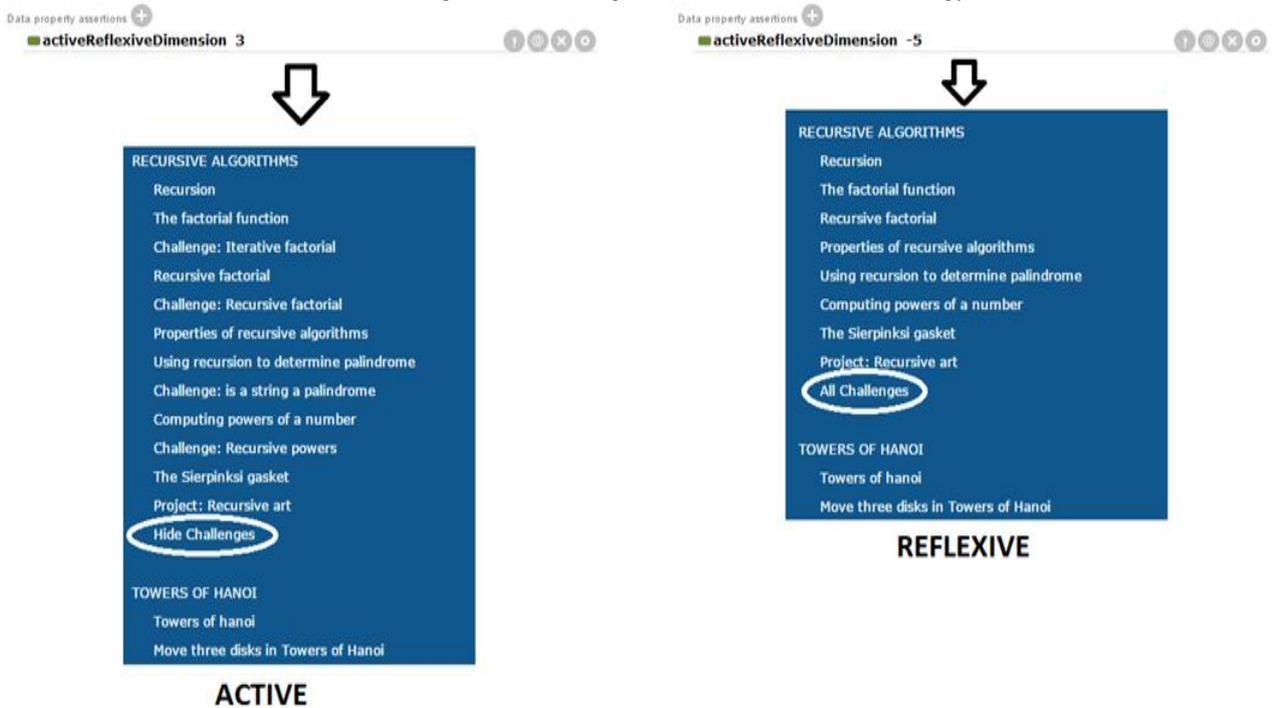

Fig.13.    Represent Active-Reflexive dimension comparison for learner's learning style

Fig. 14.is indicative of the *Sensing-Intuition dimension* of a learner's style. On the left of the illustration, sensing learners are shown. They have regular quizzes and an option to hide them, which goes towards changing the learning dimensions in the log file. Similarly, on the right, intuition learners are shown. Intuition learners have their quizzes hidden, but given with an option of *"All Quizzes",* which allowed them to take quizzes and alter the learning dimensions in the log file.

Fig. 15. is characteristic of the *Visual-Verbal dimension* of a learner's style. Visual learners are on the left of the illustration. Their primary learning style is via videos, and if applicable, with an option of *"Text Explanation"*, which when clicked enough times, modifies the learning dimensions. Verbal learners are on the right of the illustration. They learn via text with an option of watching a video, if applicable. If they opt for it, for a certain number of times, their learning dimensions will reflect the same.

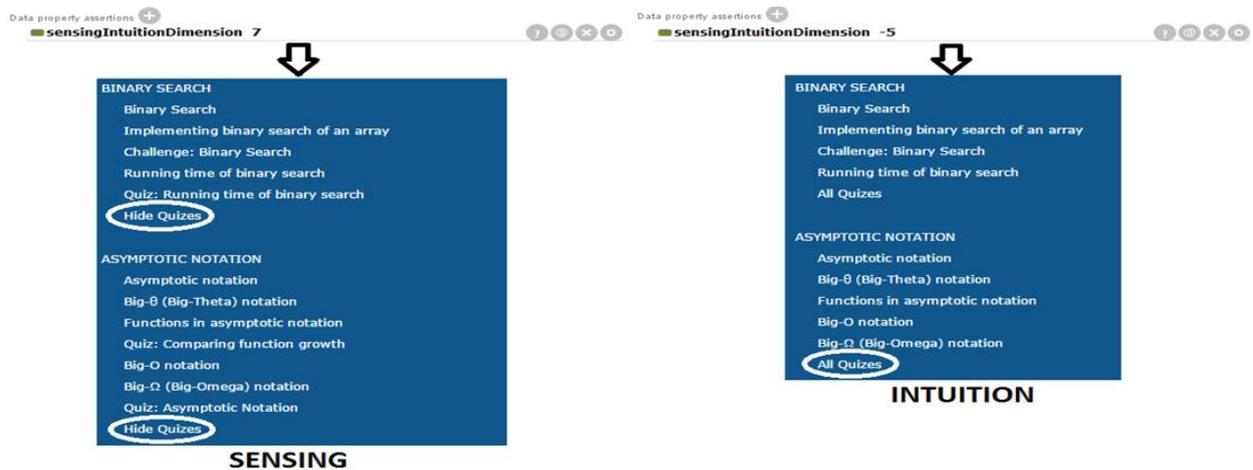

Fig.14. Represent Sensing-Intuition dimensions comparison for learner's learning style

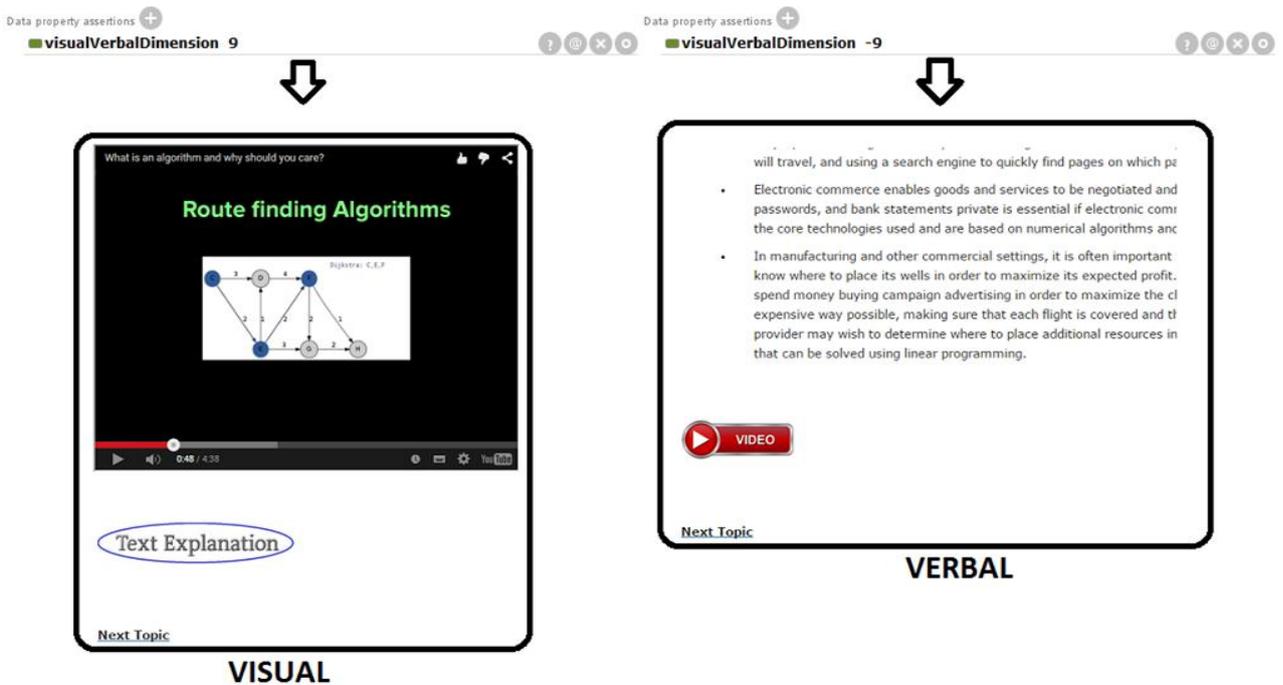

Fig.15. Represent Visual-Verbal dimensions comparison for learner's learning style

Fig. 16. is characteristic of the *Sequential-Global dimension* of a learner's style. Sequential learners on the left are presented with a content view and an option to choose a gallery view. The option has been circled in the illustration. Selection of this option goes towards modifying the learner dimensions. Global Learners, on the right, are presented with a gallery view. The global learners are provided with a content view button which has been circled in the illustration. The selection of the option goes towards changing the learning dimensions.

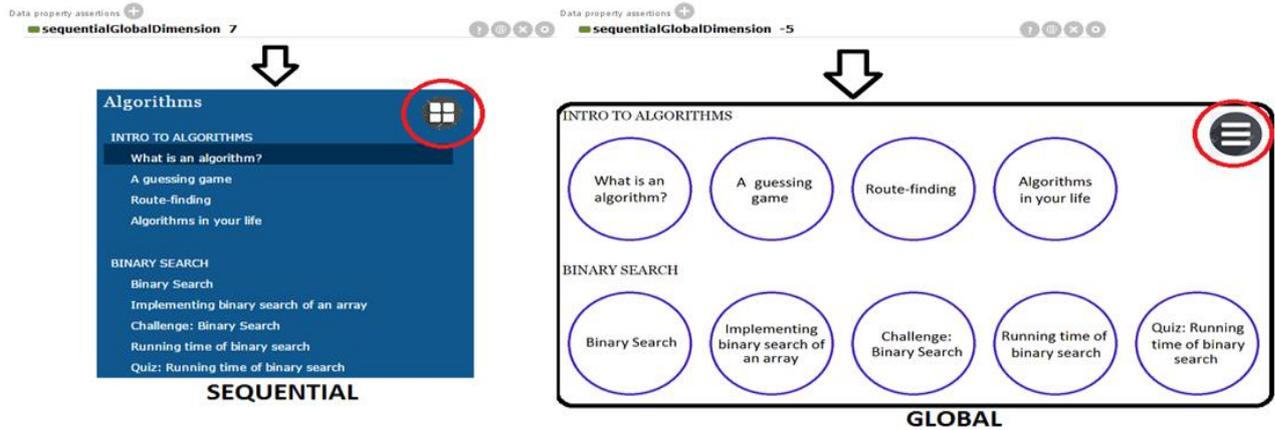

Fig.16.    Represent Sequential-Global dimensions comparison for learner's learning style

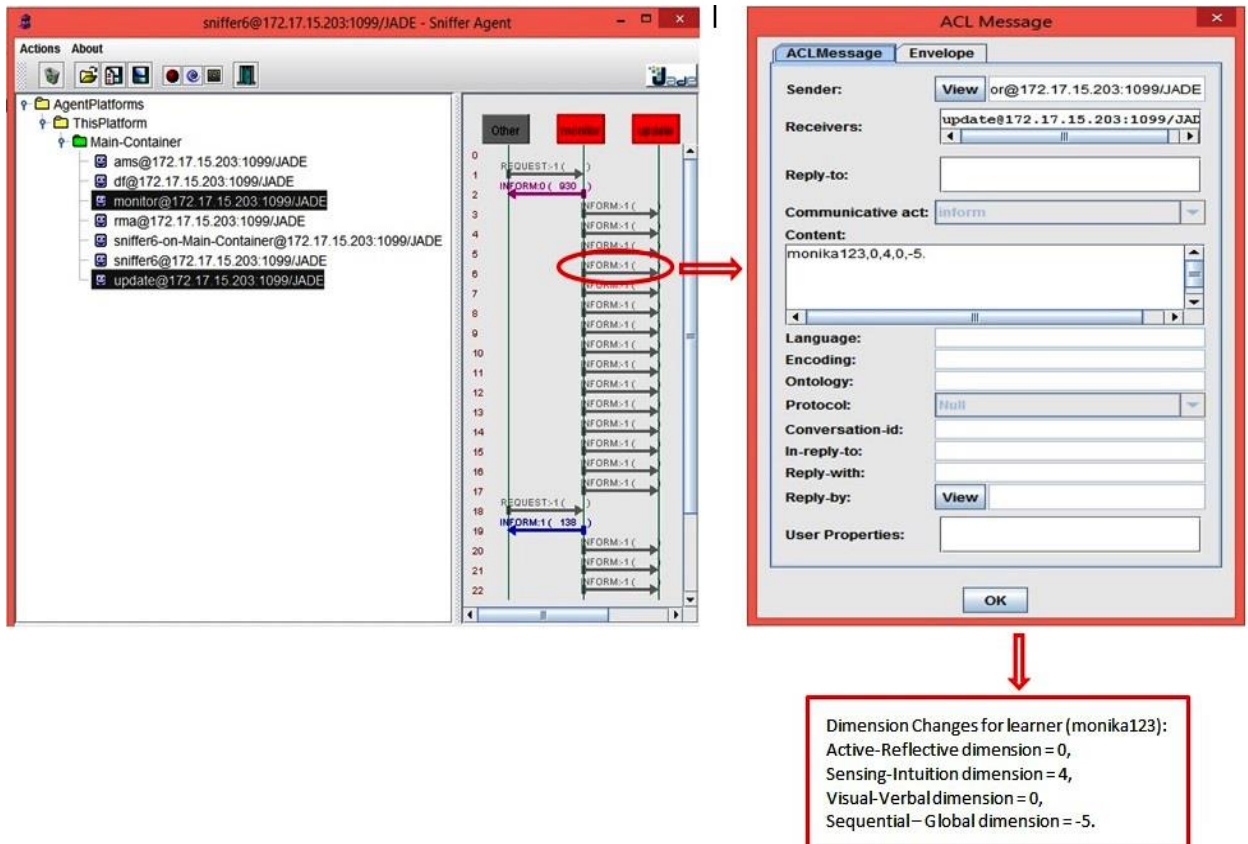

Fig.17.    Agent collaboration (Sniffer agent depicting communication between monitor and update)

**Table.1:** Describing "Active-Reflective", "Sensing-Intuition", "Visual-Verbal", "Sequential-Global" dimension changes for learner's style (monika123):

| Initial Dimension Values | Detected Change Values | Change Dimension Values | Updated Change Values |
|---|---|---|---|
| 1, 3, -1, 1 | **0, 4, 0, -5** | 1, 1, -1, 1 | 3, -3, 0, 0 |
| 1, 1, -1, 1 | -7, -6, 3, -8 | -1, -1, -1, -1 | -2, -1, 3, -3 |
| -1, -1, -1, -1 | 11, 12, 16, 18 | 3, 3, 5, 5 | 1, 2, 1, 3 |
| 3, 3, 5, 5 | 0, 0, -4, -1 | 3, 3, 5, 5 | 0, 0, -4, -1 |
| 3, 3, 5, 5 | -6, -4, -8, -7 | 1, 3, 3, 3 | -1, -4, -3, -2 |

Order of Values: "Active-Reflective", "Sensing-Intuition", "Visual-Verbal", "Sequential-Global" dimension

For updating the learner dimension *Agent:Monitoring* and *Agent:Updating* communicate with each other. The sniffer agent is used to demonstrate the following interaction. In the illustration, the *Agent:Monitoring* informs about the learning dimension changes as an ACL message. The *Agent:Updating* acknowledges the same by a message. It is indicated as a message in the sequence of message passing depicted by the sniffer agent. The ACL message informs the update agent that the dimension change has crossed a threshold in order to change them in the ontology. For example, Change in dimension for learner (monika123), Active-Reflective dimension = "0", Sensing-Intuition dimension = "4", Visual-Verbal dimension = "0", Sequential-Global dimension = "-5" of a learner's style as shown in Fig. 17.

Felder-Silverman model Questionnaires provide a quantitative method for initializing dimension values. In our proposed e-learning model we depict a Table 1 in which the results are seem to express the learner's dimension values as shown in a before and after state, each time the monitor agent inspects the ontology for change. For example, learner (monika123) dimension values change for Active-Reflective, Sensing-Intuition, Visual-Verbal and Sequential-Global dimension are 0, 4, 0, -5 respectively as shown in Fig. 17. Each time the change is noticed to be five or greater, the dimension is updated by the update agent by two to maintain an odd number. The values which are stored as data attributes in the ontology are updated since the changes have been incorporated in the actual dimension values, in order to provide adaptive and personalized learning for learners.

4.2 DISCUSSION

The discussion articulates the meaning of the results presented. As shown in Fig. 13. Active learners are regularly provided with challenges, since those learners liked to try out what they've learnt. Challenges provide a platform to exercise the knowledge. Reflexive learners on the other hand, are not regularly provided with challenges; as such learners would rather have time to think over the knowledge learnt, instead of practicing exercises based on them, immediately. Providing them with alternate options helps the system adapt to their changing behaviour and allow adaptability.

Fig. 14.shows a comparison between sensing and intuition learners. Sensing learners are very particular about the facts and details of a particular topic. It allows the learners to fully understand the topic. Intuition learners on the other hand, are more interested in theories and principles, and less concerned about the details involving the topic. Therefore, quizzes are provided for sensing learners and are hidden from intuition learners. The option of hiding and revealing the quizzes to sensing and intuition learners, respectively, helps provide alternatives and monitor their behaviour.

Visual and Verbal learners are shown in Fig. 15. Visual learners absorb better with pictures and an audio visual environment. Due to this reason they are primarily provided with a video comprising of a higher number of pictorial aids and audio explanations. They have an option to look at the text explanations as well, signalling the system a potential change in behaviour. Verbal learners are given text explanations by default as they are better learners when provided with verbal explanation rather than diagrams. Verbal learners also have an option to watch the video if they aren't completely satisfied with the explanation or for a more complete learning. If opting for a certain number of times the system alters the dimensions accordingly.

The Sequential and Global learners are shown via Fig. 16. Sequential learners are comfortable with a linear learning style in which knowledge is presented incrementally. Due to this, a content layout is provided to learners, which shows the order of learning clearly. However, if they want they can switch to a gallery view to understand the entire content as a whole. Global learners are provided the gallery view by default, because they preferred to look at the entire content as a whole to understand a given topic. If they opt to look at the order of the content they are allowed to switch to a content view. These alternatives are noticed by the log file to adapt to their changing behaviour.

As shown in Fig.17, *Agent:Update sends* the dimension changes to the *Agent:Monitor periodically* to maintain a real time adaptive system. Whenever sufficient changes are present in the log files *Agent:Monitoring* does the required activities to make the changes permanent in the ontology.

Such a proposed system has several applications in the real world. Personalization not only includes objectifying the learner's styles, but to also monitor learner's usage of the system to conform to the current learning styles. Such systems can help learners maintain focus with their changing patterns. This focus can improve productivity not only in educational institutions, but also in industrial training and learning. Integrating with the semantic web helps with the reusability of the system, providing a valid milestone for further research in the current scheme of study. Keeping the Felder-Silverman model as the foundation of the study, it helps to keep up-to-date with the latest learning styles. It can be credited to the fact that the learning model is heavily researched in statistical domains and practical situations.

Table.2: The evaluation questionnaire survey for proposed E-learning system

| Dimension | Question to ask learners |
|---|---|
| **Learner dimension** | Q1. How familiar are you with an e-learning platform?<br>Q2. The attitude towards the use of computer/laptop/mobile for e-learning purpose?<br>Q3. Your understanding about the provided e-learning course content? |
| **Instructor dimension** | Q4. How instructor organizes the content to meet learner's objectives?<br>Q5. How timely e-learning content is updated on the e-learning system, by the instructor?<br>Q6. The amount of time given for preparation of the quiz? |
| **Course dimension** | Q7. How engaging and personalized was the course content?<br>Q8. How relevant were the topics covered in the course?<br>Q9. Was the Course content direct and comprehensible? |
| **Design dimension** | Q10. The interface design of an e-learning system?<br>Q11. Quality of video, audio and text used as e-learning material?<br>Q12. Responsiveness of the e-learning system? |
| **Technology dimension** | Q13. Learning management system setup?<br>Q14. How well did the Felder-Silverman Learning Model determine your learning style?<br>Q15. Rate the e-learning platform on the basis of the following characteristics:<br>  i. Active-Reflexive dimension<br>  ii. Sensing-Intuition dimension<br>  iii. Visual-Verbal dimension<br>  iv. Sequential-Global dimension |

## 5. EVALUATION OF PROPOSED E-LEARNING SYSTEM

*5.1 Evaluation Criteria*

In 2010, Lockee et al. proposed an approach for determining quality of e-learning system called "openECBCheck" [50]. The validation of our proposed e-learning system has been done through questionnaire survey based on dimensional factors. Sun et al. [51] proposed six dimensions influencing learner's satisfaction viz., Environmental dimension: Feedback from learner in the form of questionnaire survey Learner dimension: Q1 to Q3, Instructor dimension: Q4 to Q6, Course dimension: Q7 to Q9, Design dimension: Q10 to Q12 and Technology dimension: Q13 to Q15 as shown in Table 2.

*5.2 Limitation for Questionnaires surveys for proposed e-learning system:*

Limitations for leaner and research scholar from both perspectives: Here, learner is a person who responds to the questionnaire and research scholar is a person who prepares the survey questionnaire

- Sample size is limited due to targeted population as it depends on the content provided for e-learning.
- Survey form can miss some questions that need to be answered by the learner/user.
- Learner can interpret questions in different scenarios
- Learner can interpret the question from different perspectives.
- The e-learning content store in *course.owl* will not be applicable to other domains because the proposed e-learning system will cover only computer science e-learning content for now, which can't be understood by another domain leaner/user.

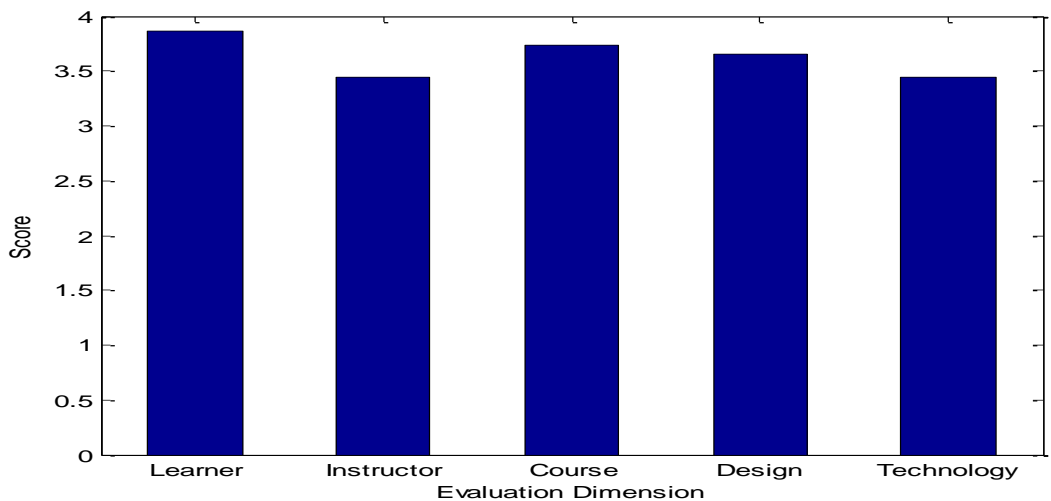

Fig.18. Average score

*5.3 The Effective Evaluation of Proposed E-learning system*

In our proposed e-learning system, we are considering feedback of questionnaire survey as Environmental dimension. The questionnaire survey consists of 15 questions with a score value of a question ranging from 1 to 5. The evaluation dimensions of our system are: Learner dimension, Instructor dimension, Course dimension, Technology dimension and Design dimension on the basis of which Average score is calculated for each dimension. These evaluation dimensions are rated on a scale of 1 to 5 points and the results are displayed in the bar chart as shown in Fig. 18. For "Learner dimension" the average score is 3.87. Similarly, "Instructor dimension" is rated as 3.45, "Course dimension" is rated as 3.74, "Technology dimension" is rated as 3.45 and "Design dimension" is rated as 3.659.

## 6. CONCLUSION & FUTURE WORK

The proposed system integrates an e-learning application with the semantic web via domain ontology. The ontologies created for the system not only provide re-usable content for the future applications with similar purposes, but also represents the concept hierarchy structure as a domain ontology with more expressive relations. The standards provided by Ontology web language (OWL) have made it trivial to understand the semantics of the knowledge base. In the proposed system, we have maintained two ontologies namely, the *course.owl and user.owl,* which are used to store learning materials and to implement the Felder-Silverman learning style model respectively. The Felder-Solomon Index of Learning Styles (ILS) is a list of questionnaires which are used to exercise the learner's learning preferences and present the information to the learner accordingly. The learning style model lays the foundation of a personalized environment, focusing on the learner's pattern.DL Query also provides an efficient way of extracting required information from the application's ontology. *HermiT* reasoner is used to determine the consistencies in *user.owl* and *course.owl* ontologies

The collaborative software agents notice to alter the learner's dimension values according to the learner's usage of the application. Contradicting learning behaviour is noticed by the agents and the preference is changed accordingly. Adaptability has been realized by deploying JADE agents, namely *Agent:Updating* and *Agent:Monitoring*. The Communication between the log file of the application and the *user.owl* via the *JadeGateway*, is the learner's changing patterns which are monitored and modified in real-time. Ontologies deployed on DigitalOcean's remote cloud host provide an expanded and secure environment to proposed e-learning system. Therefore, the final a survey was conducted to explore an adaptive, personalized e-learning application using ontology to integrate with the semantic web cloud services to, employ an incremental model and a multi-agent system to recognize adaptability in the learner's behaviour.

Future studies can be conducted to objectify the learner's personalization in a better way by considering the exact value of a particular dimension in the Felder-Solomon learning style index. Lecturers can be actualized in the system by ascertaining their teaching style by observing learners. Personalized e-learning course and its contents using adaptive Learning Path Sequence (LPS) can be recommended for learners. Evaluation parameter "Environmental dimension" can be improved by deploying proposed e-learning system on discussion forums, polling e-mail based tool, chat and Instant Messaging (IM) and audio & video conferencing. Assessing readiness of our proposed e-learning in IOE (Internet of Everything) scenario. Our proposed approach provides a motivation for advanced learning by simultaneously supporting the vision of IOE for higher education among learners. Enhancement in technology leads to the opportunities for the disabled learners by providing training in a better way. Even disabled learners can be targeted in IOE scenario for advanced digital education with various features like cloud's store's ontology, where agents interact to provide adaptive and personalized learning.